\documentclass[11pt]{article}

\usepackage{amsbsy}
\usepackage{amsmath}
\usepackage{multirow}
\usepackage{graphicx}

\begin{document}

%title
\large\textit{\textbf{Optimal B-Robust Estimation for the Parameters of Marshall-Olkin Extended Burr XII Distribution and Application for Modeling Data from Pharmacokinetics Study}}
\vspace{0.25in}

%authors
\textsuperscript{a}Ye\c{s}im G\"{u}ney, \textsuperscript{b}\c{S}enay \"{O}zdemir ,\textsuperscript{a}Yetkin Tua\c{c} and \textsuperscript{a}Olcay Arslan
\vspace{0.25in}

%address

\textsuperscript{a}Department of Statistics, Ankara University, 06100, Ankara, Turkey,\\ \textit{Contact}:ydone@ankara.edu.tr\\
\textsuperscript{b}Department of Statistics, Afyon Kocatepe University, 03200, Afyonkarahisar, Turkey.
\vspace{0.25in}

%abstract

\begin{abstract}
Marshall-Olkin Extended Burr XII (MOEBXII) distribution family, which is a generalization of Burr XII distribution proposed by Al-Saiari et al. \cite{MOEB} , is a flexible distribution that can be used in many fields such as actuarial science, economics, life testing, reliability and failure time modeling. The parameters of the MOEBXII distribution are usually estimated by the maximum likelihood (ML) and least squares (LS) estimation methods. However, these estimators are not robust to the outliers which are often encountered in practice. There are two main purposes of this paper. The first one is to find the robust estimators for the parameters of the MOEBXII distribution. The second one is to use this distribution for modeling data from pharmacokinetics study. To obtain the robust estimators we use the optimal B robust estimator proposed by Hampel et al.  \cite{Hampel}. We provide a simulation study to show the performance of the proposed estimators for the ML, LS and robust M estimators. We also give a real data example to illustrate the modeling capacity of the MOEBXII distribution for data from pharmacokinetics study.
\end{abstract}

\vspace{0.25in}

\textit{\textbf{Class codes:}}
\vspace{0.25in}

\section{Introduction}
\label{sec:intro}

 Burr \cite{burr} introduced a family of continuous distributions that includes twelve types of cumulative distribution functions with different shapes. Since then, Burr XII has attracted the most attention in many different  fields \cite{actuarial,economics,rel1,rel3,rel2,Lawless,rel4}.
  The Burr distribution has relationship with several distributions and some of them summarized by  Rodriguez  \cite{summary2} and Tadikamalla \cite{summary}.     For the purpose of providing greater flexibility in modeling data, many generalizations of the Burr XII distribution have been introduced in literature. One of these generalizations is based on the Marshall-Olkin transformation which provide more flexible distribution Marshall and Olkin \cite{MO}. Marshall and Olkin introduced a method of obtaining a family of distributions by introducing an additional parameter $\alpha$. $F(x)$ and $\bar{F}(x)=1-F(x)$ are  cumulative distribution function (cdf) and the survival function of the baseline distribution, respectively. Then the related Marshall-Olkin (MO) extended distribution has a survival function defined as follows

\begin{equation}\label{1}
\bar{F}_{\alpha}(x)=\frac{\alpha\bar{F}(x)}{1-\bar{\alpha}\bar{F}(x)}
\end{equation}

\noindent where $\alpha>0$ is an additional parameter and $\bar{\alpha}=1-\alpha$. One can obtain the initial family as a particular case with $\alpha=1$. By using the transformation given in (\ref{1}) many different distribution generalizations are defined in the literature. One of them is introduced by \cite{MOEB} which is called the Marshall-Olkin extended Burr type XII (MOEBXII) distribution.

    Several researchers have considered the parameter estimation methods for the Burr XII distribution. For instance parameter estimation of the Burr XII distribution with the ML method to life test data has been considered by Wingo \cite{ML,ML2}. The minimum variance linear unbiased estimators (MVLUE), the best linear invariant estimators (BLIE) and the ML estimators based on n-selected generalized order statistics are presented for the parameters of the Burr XII distribution by Malinowska et al. \cite{order}. Shao \cite{Shao} gives a complete investigation on the behaviors of the ML estimates based on uncensored and right-censored data. Wang and Cheng \cite{Wang} present a robust regression method to estimate the parameters of the Burr XII distribution. Do\u{g}ru and Arslan \cite{rburr,bburr} proposed an estimator based on M estimation and optimal B-robust (OBR) estimator to estimate the parameters of Burr XII distribution. However, a few estimation procedures have been proposed for the parameters of the MOEBXII distribution in the literature. For example, ML and Bayes estimators of the parameters of the MOEBXII distribution have been derived by  Al-Saiari et al.  \cite{MOEB}. Recently, G\"{u}ney and Arslan  \cite{yesim} considered the MOEBXII distribution and robust parameter estimation methods. Although the ML estimators have good properties, considering the data contains outliers, it is known that the ML estimators do not perform well. In this paper, we propose to use the OBR estimator to estimate the parameters of the MOEBXII distribution.

    The pharmacokinetics properties of the drug are among the most important drug characteristics for optimal treatment after the selection of the appropriate drug in the treatment of a disease. The most appropriate daily dose to achieve the effective plasma level is determined by these properties. Among these properties one of the most important pharmacokinetics property is plasma drug concentration. The maximum concentration $(C_{max})$ and the time taken to reach the maximum concentration $(T_{max})$ are also important variables for the pharmacokinetics studies. These variables can be easily estimated after obtaining data fit. However, to obtain the reliable estimates of $C_{max}$ and $T_{max}$, trustfully modeling of the plasma drug concentration is necessary (For more details see \cite{pharmaco}).

    In summary, the aim of this paper is twofold. First to use the OBR estimation method to obtain  the robust estimators of the parameters of the MOEBXII distribution. By doing this, we will gain robustness against to outliers. We have already mentioned that this distribution can be used for modeling data from several different areas, however not for the pharmacokinetics data. The second aim of this work is to use the MOEBXII distribution for modeling pharmacokinetics data with the robust estimators.

    The remainder of the paper is organized as follows; In Section ~\ref{sec:1}, we describe the MOEBXII distribution. In Section ~\ref{sec:2}, we briefly describe the ML, LS, robust M estimation method and we give the OBR estimation method for the parameters of MOEBXII distribution. In section ~\ref{sec:7}, simulation study is presented to compare the performance of the OBR estimation method with that of ML, LS and robust M estimation methods. In Section ~\ref{sec:8}, we analyze a data set from a pharmacokinetics study. Finally, conclusions are given in section ~\ref{sec:9}.

\section{Marshall-Olkin Extended Burr XII Distribution}
\label{sec:1}
    The probability density function (pdf) and the cdf of Burr XII distribution is given respectively
\begin{equation}\label{2}
	f(x;c,k)=ck\frac{x^{(c-1)}}{\left(1+x^{c}\right)^{k+1}}, x\geq0,
\end{equation}

\begin{equation}\label{3}
	F(x;c,k)=1-\frac{1}{\left(1+x^{c}\right)^{k}},x\geq0
\end{equation}

\noindent where $c$ and $k>0$ are the shape parameters. Substituting (\ref{3}) in (\ref{1}) we obtain the Marshall-Olkin Extended Burr XII distribution denoted by MOEBXII$(\alpha,c,k)$ with the following pdf and cdf respectively

\begin{equation}\label{4}	
f(x;\alpha,c,k)=\alpha ck
\frac{x^{(c-1)}\left(1+x^{c}\right)^{-(k+1)}}{\left[1-(1-\alpha)\left(1+x^{c}\right)^{-k}\right]^{2}}, x\geq0,
\end{equation}

\begin{equation}\label{5}	 F(x;\alpha,c,k)=\frac{1-\left(1+x^{c}\right)^{-k}}{1-(1-\alpha)\left(1+x^{c}\right)^{-k}},x\geq0
\end{equation}

\noindent where $\alpha$,$c$ and $k>0$ Al-Saiari et al. \cite{MOEB}. For $\alpha=1$ it corresponds to the Burr XII distribution with two parameters $c$ and $k$. The MOEBXII distribution contains distributions with various shapes such as bell-shaped, right-skewed, L-shaped. Therefore, it has a crucial advantage of flexibility to fit datasets with various shapes. For more details, see  \cite{MOEB}.

\section{Estimation of the Parameters of MOEBXII Distribution}
\label{sec:2}

In this section, we consider the ML, LS, robust M and OBR estimation methods to estimate the parameters of the MOEBXII distribution.

\subsection{Maximum Likelihood Estimation}
\label{sec:3}

    Let $x=(x_{1},x_{2},\ldots,x_{n})$ be a random sample of size n from MOEBXII$(\alpha,c,k)$. Then, the log-likelihood function based on the given random sample is

\begin{equation}\label{6}
\begin{array}{lll}
  l(\alpha,c,k) & = & nlog(\alpha ck)+(c-1)\sum_{i=1}^{n}{logx_{i}}\\
  & - &(k+1)\sum_{i=1}^{n}{log(1+x_{i}^{c})}  \\
  & - &2\sum_{i=1}^{n} log(1-(1-\alpha)\left(1+x_{i}^{c}\right)^{-k}).
\end{array}
\end{equation}

ML estimations of the parameters, denoted by $\alpha$, $c$ and $k$, can be obtained as the simultaneous solutions of
\begin{equation}\label{7}
	\frac{\partial l}{\partial \alpha} = \frac{n}{\alpha}
     - 2\sum_{i=1}^{n}{\frac{\left(1+x_{i}^{c}\right)^{-k}}{1-(1-\alpha)\left(1+x_{i}^{c}\right)^{-k}}}=0,
\end{equation}

\begin{equation}\label{8}
\begin{array}{lll}
	\frac{\partial l}{\partial c} &=& \frac{n}{c}+\sum_{i=1}^{n}{log x_{i}}-(k+1)\sum_{i=1}^{n}\frac{x_{i}^{c}log(x_{i})}{\left(1+x_{i}^{c}\right)}\\
&-&2k(1-\alpha)\sum_{i=1}^{n}{\frac{x_{i}^{c}(1+x_{i}^{c})^{-(k+1)}log(x_{i})}{(1-(1-\alpha)(1+x_{i}^{c})^{-k})}}=0,
\end{array}
\end{equation}

\begin{equation}\label{9}
\begin{array}{lll}
	\frac{\partial l}{\partial k}& = & \frac{n}{k}-\sum_{i=1}^{n}log(1+x_{i}^{c})-(k+1)\sum_{i=1}^{n}\frac{x_{i}^{c}log(x_{i})}{(1+x_{i}^{c})}\\
& - &2(1-\alpha)\sum_{i=1}^{n}\frac{(1+x_{i}^{c})^{-k}log(1+x_{i}^{c}}{1-(1-\alpha)(1+x_{i}^{c})^{-k}}=0.
\end{array}
\end{equation}

    Since the equations (\ref{7}), (\ref{8}) and (\ref{9}) cannot be solved analytically, we can obtain the estimates of the parameters of interest b
y using the numerical methods.

\subsection{Least Squares Estimation}
\label{sec:4}
LS estimation method was first suggested by  Swain et al. \cite{Swain} to estimate the parameters of the beta distribution. It has been also used for the parameters of the Burr distribution Hossain and Nath \cite{outlier} and MOEBXII distribution G\"{u}ney and Arslan \cite{yesim}. The LS estimation method is based on minimizing the following function

\begin{equation}\label{10}
    \begin{array}{lll}
     S(\alpha,c,k) &=& \sum_{i=1}^{n}\left(\widehat{F}(x_{i})-F(x_{i})\right)^{2}\\
                &=& \sum_{i=1}^{n}\left(\widehat{F}(x_{i})-\frac{1-(1+x_{i}^{c})^{-k}}{1-(1-\alpha)(1+x_{i}^{c})^{-k}}\right)^{2}.
    \end{array}
\end{equation}

    Since the cdf of the MOEBXII distribution is a non-linear function, the minimization of equation (\ref{10}) is difficult.  To handle this problem  following equation can be used.

    \begin{equation*}
    \log\left(\frac{1}{1-F(x)}\right).
    \end{equation*}

Let define $y_{(i)}=log\left(\frac{1}{1-\widehat{F}_{X_{(i)}}(x)}\right)$ and $u_{(i)}=log\left( \frac{1}{1-F\left( x_{(i)}\right) }\right) $ where

\begin{equation}\label{11}
    \widehat{F}_{X_{(i)}}(x)=\frac{i-0.5}{n},  i=1,2,\ldots,n
\end{equation}

and $X_{(i)}$ denotes the $i.$ order statistics of the random sample from MOEBXII. Thus, the LS estimates of the parameters can be obtained by minimizing following objective function:

\begin{equation}\label{12}
    S(\alpha,c,k)=\sum_{i=1}^{n}{\left(y_{(i)}-u_{(i)}\right)^{2}}
\end{equation}

\noindent where $y_{(i)}=log\left(\frac{1}{1-F_{X_{(i)}}(x)}\right)$ and $u_{(i)}=\left( \frac{1}{1-F\left( x_{(i)}\right) }\right) $. To obtain the LS estimates, the following equations should be solved with respect to $\alpha$,$c$ and $k$.

\begin{equation}\label{13}
\sum_{i=1}^{n}\left( y_{(i)}-u_{(i)}\right) \left( \frac{%
1-(1+x_{(i)}^{c})^{-k}}{\alpha \left[ 1-\left( 1-\alpha \right)
(1+x_{(i)}^{c})^{-k}\right] }\right) =0,
\end{equation}

\begin{equation}\label{14}
\sum_{i=1}^{n}\left( \left( y_{(i)}-u_{(i)}\right) \frac{%
x_{(i)}^{c}\log \left( x_{(i)}\right) \left( 1-(1+x_{(i)}^{c})^{-k}\right) }{%
(1+x_{(i)}^{c})^{-k}\left[ 1-\left( 1-\alpha \right) (1+x_{(i)}^{c})^{-k}%
\right] }\right) =0,
\end{equation}	

\begin{equation}\label{15}
\sum_{i=1}^{n}\left( \left( y_{(i)}-u_{(i)}\right) \frac{\log
(1+x_{(i)}^{c})}{\left[ 1-\left( 1-\alpha \right) (1+x_{(i)}^{c})^{-k}\right]
}\right) =0.
\end{equation}

\subsection{M Estimation}
\label{sec:5}
M estimation method was first revealed by Huber et al.\cite{hub}. G\"{u}ney and Arslan \cite{yesim} have been proposed a parameter estimation method based on M estimation for the parameters of MOEBXII. The method based on the minimizing the following objective function with respect to the parameters of interest.

\begin{equation}\label{16}
    Q(\alpha ,c,k)=\sum_{i=1}^{n}\rho \left( y_{i}-u_{i}\right).
\end{equation}	

 $\rho$ function in (\ref{16}) is more resistant than the square function to the outliers in data. It is also non-negative, symmetric function and $\rho(0)=0$.

 The estimates based on M estimators can be obtained by solving the following non-linear equations based on the derivatives of objective function (\ref{16}).

%\begin{small}
%\begin{equation}\label{21}
%\widehat{k}=\frac{\sum_{i=1}^{n}\omega _{i}\left( y_{i}+\log \left(
%\alpha \right) -\log \left( 1-\left( 1-\alpha \right)
%(1+x_{i}^{c})^{-k}\right) \right) \frac{\log (1+x_{i}^{c})}{\left[ 1-\left(
%1-\alpha \right) (1+x_{i}^{c})^{-k}\right] }}{\sum_{i=1}^{n}\omega
%_{i}\frac{\left( \log (1+x_{i}^{c})\right) ^{2}}{\left[ 1-\left( 1-\alpha
%\right) (1+x_{i}^{c})^{-k}\right] }},
%\end{equation}
%\end{small}

%\begin{small}
\begin{equation}\label{21}
% \begin{array}{l}
\widehat{k}=\frac{\sum_{i=1}^{n}\omega _{i}\left( y_{i}+\log \left(
\alpha \right) -\log h_{i} \right)
\frac{\log (1+x_{i}^{c})}{h_{i} }}{\sum_{i=1}^{n}\omega
_{i}\frac{\left( \log (1+x_{i}^{c})\right) ^{2}}{h_{i} }},
%\end{array}
\end{equation}
%\end{small}

%\begin{equation}\label{22}
%\sum_{i=1}^{n}\left( \omega _{i}\left( y_{i}-u_{i}\right) \frac{%
%x_{i}^{c}\log \left( x_{i}\right) \left( 1-(1+x_{i}^{c})^{-k}\right) }{%
%(1+x_{i}^{c})^{-k}\left[ 1-\left( 1-\alpha \right) (1+x_{i}^{c})^{-k}\right]
%}\right) =0,
%\end{equation}	
\begin{equation}\label{22}
\sum_{i=1}^{n}\left( \omega _{i}\left( y_{i}-u_{i}\right) \frac{%
x_{i}^{c}\log \left( x_{i}\right) \left( 1-(1+x_{i}^{c})^{-k}\right) }{%
(1+x_{i}^{c})^{-k}h_{i}
}\right) =0,
\end{equation}	

%\begin{small}
%\begin{equation}\label{23}
%\log \widehat{\alpha }=\frac{\sum_{i=1}^{n}\omega _{i}\left(
%y_{i}-k\log (1+x_{i}^{c})-\log \left( 1-\left( 1-\alpha \right)
%(1+x_{i}^{c})^{-k}\right) \right) \left( \frac{1-(1+x_{i}^{c})^{-k}}{\alpha %
%\left[ 1-\left( 1-\alpha \right) (1+x_{i}^{c})^{-k}\right] }\right) }{%
%\sum_{i=1}^{n}\omega _{i}\left( \frac{1-(1+x_{i}^{c})^{-k}}{\alpha %
%\left[ 1-\left( 1-\alpha \right) (1+x_{i}^{c})^{-k}\right] }\right) }
%\end{equation}
%\end{small}
\begin{small}
\begin{equation}\label{23}
\log \widehat{\alpha }=\frac{\sum_{i=1}^{n}\omega _{i}\left(
y_{i}-k\log (1+x_{i}^{c})-\log h_{i} \right) \left( \frac{1-(1+x_{i}^{c})^{-k}}{\alpha %
h_{i} }\right)}{%
\sum_{i=1}^{n}\omega _{i}\left( \frac{1-(1+x_{i}^{c})^{-k}}{\alpha %
h_{i} }\right) }
\end{equation}
\end{small}

\noindent where
$h_{i}=\left[ 1-\left(1-\alpha \right) (1+x_{i}^{c})^{-k}\right]$
and $\omega_{i}$'s are the weights. We use following  Tukey's $\rho$ function
\begin{equation}\label{20}
\rho=\left\{
         \begin{array}{ccc}
           1-(1-(x/b)^2)^{2} & , & x\leq b \\
           1 & , & x > b \\
         \end{array}
       \right.
\end{equation}

and the weights are

\begin{equation}\label{24}
\omega_{i}=\left(\left(1-\left(\frac{(y_{i}-u_{i})}{(b)}\right)^{2}\right)^{2}I(|(y_{i}-u_{i})| \leq b)\right).
\end{equation}

Here $b$ is called the robustness tuning constant.

\subsection{Optimal B-Robust Estimation}
\label{sec:6}

    The class of the OBR estimators was defined by Hampel et al. \cite{Hampel}. The OBR estimation method is a robust alternative modification of M estimation method with bounded influence function. It is also the most efficient one in the class of robust M-estimators. In literature,Victoria-Feser  \cite{Vic1} and  Victoria-Feser and Ronchetti \cite{Vic2} introduced the OBR estimators to estimate the parameters of the Pareto and the gamma distributions. Dogru and Arslan\cite{bburr} introduced the OBR estimation method for the Burr XII distribution. Dogru et al. \cite{bhnormal} also proposed robust estimators by using the OBR estimation method for the parameters of the generalized half-normal distribution.

    According to Hampel et al. \cite{Hampel} there are two ways of defining the optimal B-robust estimation. The first one is the minimax approach defined by Huber et al. \cite{hub}. The second one is called the infinitesimal approach defined by Hampel et al. \cite{Hampel}. In this paper we will use the second approach tries to find M-estimators with bounded influence function (IF), that have minimum asymptotic variance.
    The IF can be defined in the following way. For a sample of n observations, $\underline{x}=(x_{1},x_{2},...,x_{n})$, the empirical distribution function $F_{n}(x)$ is

\begin{equation}
    F_{n}(x)=\frac{1}{n}\sum_{i=1}^{n}\delta_{x_{i}}(x)
\end{equation}

\noindent where $\delta_{x_{i}}$ denotes a point mass in $x$.

    For a parametric model $\{F_{\theta}:\theta\in\Theta\subset R^{p}\}$, estimators of $\theta$; $T_{n}$ can be represented as a statistical functional of the empirical distribution, i.e. $T_{n}(x_{1},x_{2},...,x_{n})=T_{n}(F_{n})$. Then the IF of the estimator $T_{n}$ at $F$ is given by
	
\begin{equation}
IF(x,T_{n},F_{\theta})=lim_{\epsilon\rightarrow0}\frac{T_{n}((1-\epsilon)F_{\theta}+\epsilon\delta_{x})-T_{n}(F_{\theta})}{\epsilon}.
\end{equation}

    The IF describes the relative influence of individual observations toward the value of an estimate Hampel et al. \cite{Hampel}. When the IF is unbounded, an outlier can have an overriding influence on the estimate.
    The IF of the maximum likelihood estimator is

\begin{equation}
IF=J(\theta)^{-1}s(x,\theta)
\end{equation}
	
\noindent where $J(\theta)$ is the Fisher information matrix and $s(x,\theta)=\frac{\partial}{\partial\theta}log f(x,\theta)$ is the score function. It is clear that the IF of ML estimator is not bounded if the score function is not bounded.

\begin{figure}
\centering
\includegraphics[scale=0.5]{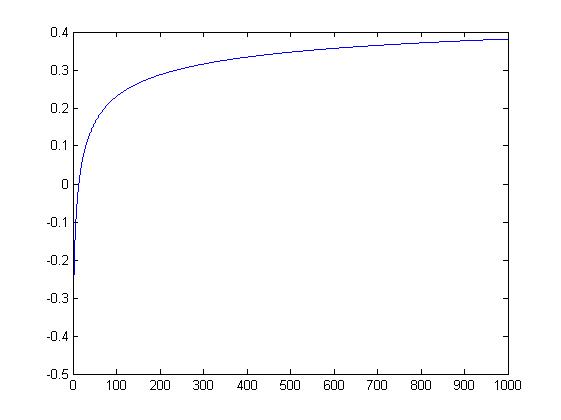}
\caption{Plot of the score function for $\alpha$ parameter with $\alpha=2$, $c=0.8$ and $k=0.5$.}
\label{Figure2}
\end{figure}

\begin{figure}
\centering
\includegraphics[scale=0.5]{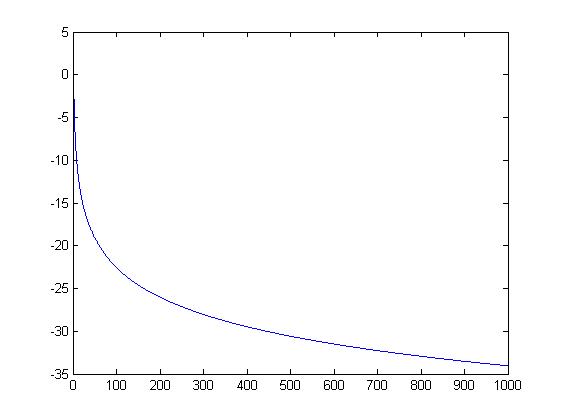}
\caption{Plot of the score function for $c$ parameter with $\alpha=2$, $c=2$ and $k=5$.}
\label{Figure3}
\end{figure}

\begin{figure}
\centering
\includegraphics[scale=0.5]{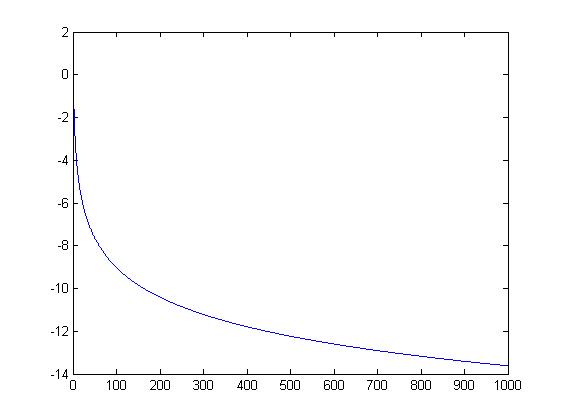}
\caption{Plot of the score function for $k$ parameter with $\alpha=2$, $c=2$ and $k=5$.}
\label{Figure4}
\end{figure}

    Considering the elements of score vector for the MOEBXII distribution, (\ref{7})-(\ref{9}), one can observe that the score functions for $\alpha$, $c$ and $k$ are unbounded function of $x$ as in the Burr XII distribution case. This can be easily checked from the Figure \ref{Figure2},\ref{Figure3} and \ref{Figure4}.

If $\alpha$, $c$ and $k$ are estimated by using ML and LS, these estimators may suffer from outliers. We propose to use the OBR estimation method instead of ML and LS estimation in the presence of outliers.

    We consider the standardized OBR estimators defined as the solution with respect to $\theta$ of
	
\begin{equation}\label{25}
%\sum_{i=1}^{n}\Psi(A(\theta)(s(\theta,x_{i})-a(\theta)))=\sum_{i=1}^{n} W(\theta,x_{i},c_{B})(s(\theta,x_{i})-a(\theta))=0
\begin{array}{ll}
 &\sum_{i=1}^{n}\Psi(A(\theta)(s(\theta,x_{i})-a(\theta))) \\
 =& \sum_{i=1}^{n} W(\theta,x_{i},c_{B})(s(\theta,x_{i})-a(\theta)) \\
 =& 0
\end{array}
\end{equation}

where

\begin{equation}\label{26}
W(\theta,x_{i},c)=\min\left(1,\frac{c^{*}}{\| A(\theta)(s(\theta,x_{i})-a(\theta))\|}\right).
\end{equation}	

Here $c_{B}$ is a tuning parameter, $c_{B}\geq \sqrt{dim(\theta)}, \parallel\cdot\parallel$ denoted the Euclidean norm, $s(\cdot)$ is the score function, $A(\theta)$ is a $dim(\theta)\times dim(\theta)$ scaling matrix and $a(\theta)$ is a $dim(\theta)$ centering vector determined by

\begin{equation}\label{27}
E\left[\Psi(\theta,x)\Psi(\theta,x)^{T}\right]= \left[A(\theta)^{T}A(\theta)\right]^{-1},
\end{equation}
\\	
\begin{equation}\label{28}
    E\left[\Psi(s(\theta,x)-a(\theta))\right]=0.
\end{equation}

    The OBR estimator keeps a level of efficiency close to the ML estimator because of the score function. The constant $c_{B}$, robustness constant, is typically fixed by setting the amount of efficiency loss and a bound on the IF. For higher values of $c_{B}$ the estimates gain efficiency, but lose robustness and vice versa. If the bound on the IF is removed, i.e, choose $c_{B}=\infty$  the OBR estimation method reduce to the ML estimation method.
    To compute the OBR estimates of the parameters, we follow an algorithm proposed by  Victoria-Feser and Ronchetti \cite{Vic2}.\\

      \textit{ OBRE Algorithm:}\\

\begin{enumerate}
    \item Fix precision threshold $\eta$ and initial value for $\theta_{(0)}$ (we can take the ML estimates as initial parameter values).

    Take initial values $a=0$, and $A=\left(\left[J^{-1}\right]^{T}\right)^{1/2}$ where
\begin{equation}
    J=\int s(\theta,x)s(\theta,x)^{T}dF_{\theta}(x)
\end{equation}

    is the Fisher Information Matrix.

    \item Solve the following equations with respect to $a$ and $A$
\begin{equation}
    A^{T}A=M_{2}^{-1}
\end{equation}
	
\begin{equation}
    a=\frac{\int W(\theta,x,c_{B})s(\theta,x)dF_{\theta}(x)}{\int W(\theta,x,c_{B})dF_{\theta}(x)}
\end{equation}
	
    where
\begin{small}
\begin{equation}
% \nonumber to remove numbering (before each equation)
 \begin{array}{l}
  M_{k} = \int W(\theta,x,c_{B})^{k}\left[s(\theta,x)-a(\theta)\right] \left[s(\theta,x)-a(\theta)\right]^{T}dF_{\theta}(x), \\
  k=1,2.
 \end{array}
\end{equation}
%\begin{equation}
%    M_{k}= \int W(\theta,x,c_{B})^{k}\left[s(\theta,x)-a(\theta)\right] \left[s(\theta,x)-a(\theta)\right]^{T}dF_{\theta}(x), k=1,2.
% %    M_{k}=\int W(\theta,x,c_{B})^{k}\[s(\theta,x)-a(\theta)\]\[s(\theta,x)-a(\theta)\]^{T}dF_{\theta}(x),  k=1,2.
%\end{equation}
\end{small}	

The current values of $\theta$, $a$ and $A$ are used as initial values to solve the given equations.

    \item Now compute $M_{1}$ and
\begin{equation}
    \Delta\theta=M_{1}^{-1}\left(\frac{1}{n}\sum_{i=1}^{n}W(\theta,x_{i},c_{B})^{k}\left[s(\theta,x_{i})-a(\theta)\right]\right).
\end{equation}
	 %$\Delta\theta=M_{1}^{-1}\left(\frac{1}{n}\sum_{i=1}^{n}W(\theta,x_{i},c_{B})^{k}\[s(\theta,x_{i})-a(\theta)\]\right)$.
    \item If $\|\Delta\theta\|>\nu$ then $\theta\rightarrow\theta+\Delta\theta$ and return to step 2, otherwise terminate the algorithm.
\end{enumerate}

\section{Simulation Study}
\label{sec:7}
    A Monte Carlo simulation study was conducted based on various scenarios for the number of observations, the number of outliers to examine the performance of the estimation methods; the ML, LS, robust M estimation with Tukey and OBR estimation methods. The superiority of the estimates was compared by employing the performance measures biases and Root-mean-square error (RMSE)

\begin{equation}\label{29}
    Bias\left(\hat{\theta}\right)=\frac{1}{N}\sum_{i=1}^{n}\left(\hat{\theta}_{i}-\theta\right),
\end{equation}
	
\begin{equation}\label{30}
    RMSE\left(\hat{\theta}\right)=\sqrt{\frac{1}{N}\sum_{i=1}^{n}\left(\hat{\theta}_{i}-\theta\right)^{2}},
\end{equation}
	
respectively.

    We generated $N=100$ replication from different samples sizes ($n=25$, $n=50$, and $n=100$) from the MOEBXII distribution as well as taking the different combination of parameter values $(\alpha,c,k)=(3,1,1)$, $(3,1,2)$, $(3,2,1)$, $(3,2,2)$, $(3,3,3)$, $(5,1,1)$, $(5,1,2)$, $(5,2,1)$ and $(5,2,2)$. (One can find the generating data set from the MOEBXII distribution in \cite{yesim}. In this study, outliers were generated by replacing the last values of the sample with 5$\times$largest observations.

    We choose tuning constant $b=1.345$ for Tukey's $\rho$ function to obtain M estimation. For the OBR estimation method, robustness parameter $c_{B}$ and precision threshold $\nu$ were taken as $3$ and $10^{-6}$ respectively.

    The obtained results are reported as the bias and RMSE of the methods described in Section 3 in Tables\ref{T1}-\ref{T6}.
    %\ref{T1},\ref{T2},\ref{T3},\ref{T4},\ref{T5},\ref{T6}.

    The Tables \ref{T1}-\ref{T3} presents the results from without outliers case. From Table \ref{T1}-\ref{T3}, we can observe that all the methods considered are preferable to estimate the parameters if there are no outlier in the data. It is obvious from Table \ref{T1} that the OBR estimation method shows the best performance in terms of RMSE for all parameters for the small sample size ($n=25$). The biases of the OBR estimates are lower than that of other methods for most of the values of the parameters. According to Table \ref{T2} and Table \ref{T3}, as the sample size increases, the MLE becomes the best method among the others in terms of RMSE as expected. The number of the values of the parameters that the biases of the MLE are lower than that of the others are increasing with the increasing sample size.

        We repeat the simulation for the same scenarios with outliers and the results are summarized in Table \ref{T4}-\ref{T7}. For the sample size n=25, there is one outlier, for the sample sizes n=50 and n=100, there are two and four outliers, respectively. In Table \ref{T7}, the results obtained from the data sets include more outliers than other cases.

    Table \ref{T4} shows the simulation results for the sample size n=25 with one outlier. We observe that outlier induces a large influence on the bias and RMSE of the MLE and LS whereas it has a smaller impact on the robust estimation methods. If the M and the OBR estimation methods are compared with each other, the OBR estimation method has superiority than the M estimation method in general in terms of the RMSE.

    Table \ref{T5} shows the simulation results with 2 outliers in a sample size of 50. When the outliers are put in the data, the ML and LS estimators are drastically worsen which is reflected to the higher RMSE and biases. However, the M and the OBR estimators still have undoubtedly better performance with outliers.

    Table \ref{T6} represents the simulation results with 4 outliers in 100 sample sizes. According to Table \ref{T6}, the OBR estimation method outperforms in terms of bias and standard error for the most values of the parameters among the others.

    As indicated by Table \ref{T4}-\ref{T6}, the same results are further improved in Table \ref{T7}. To sum up, we observe that the amount of efficiency we lose under the OBR estimation method is negligible in comparison to the other estimation methods in most of the cases.

    Briefly, when there are potential outliers in the data the OBR estimation method outperforms among the others in terms of the biases and RMSE.

\section{Real Data Example}
\label{sec:8}
    In this section the application of the MOEBXII distribution to a real data set is discussed to illustrate the performance of the proposed parameter estimation method. We use a data set from a pharmacy study of  Canaparo et al. \cite{data}. The data is related to the ibuprofen which is widely available as an over-the-counter treatment for pain and fever. It represents the mean plasma concentration--time profile of Ibuprofen (S) in all healthy subjects after a single 400 mg oral dose of racemic Ibuprofen. Ibuprofen blood plasma levels were computed at various time points using data from pharmacokinetics trials.

    We use the MOEBXII distribution to fit the data. The sample size is $n=65$. The estimates obtained from the ML, LS, robust M based on Tukey and OBR estimation methods are given in Table \ref{T8}. To obtain the OBR estimates of the parameters we use the following steps:\\

\begin{enumerate}
    \item[(i)] Obtain the ML estimate.
    \item[(ii)] Take $c_{B}=3$, ML estimate as an initial estimate and calculate the OBR estimate.
    \item[(iii)] Take $c_{B}=3$, the OBR estimate obtained in step (ii) as a new initial estimate and calculate the OBR estimate again (\cite{Hampel}).
\end{enumerate}

     Note that one can see \cite{Hampel} and \cite{Vic2} for further details about the selection of the robustness tuning constant (\cite{bhnormal}).\\

\begin{figure}
\centering
\includegraphics[scale=0.5]{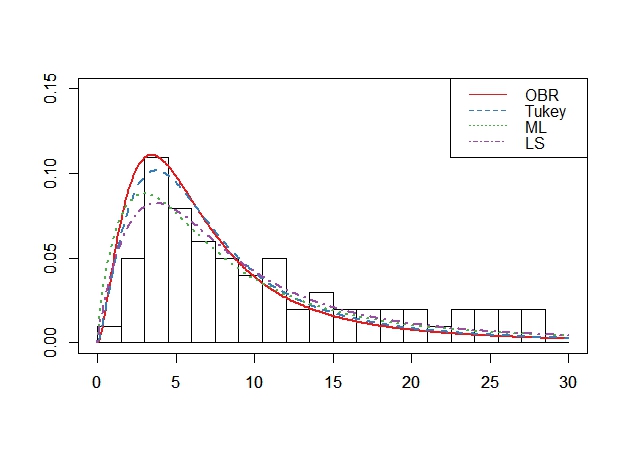}
\caption{Histogram of Ibuprofen data and the fitted densities with the ML, LS, robust M and OBR estimates}
\label{Figure5}
\end{figure}

    The fitted densities obtained from the ML, LS, robust M and OBR estimates and histogram of the ibuprofen data set are given in Figure \ref{Figure5}.

%	Figure:	[Histogram of Ibuprofen data and the fitted densities with the ML, LS, robust M and OBR estimates]

    From Figure \ref{Figure5}, it can be observed that the MOEBXII distribution is suitable to model the mean plasma concentration of ibuprofen. All of the proposed estimators are in good agreement in terms of fitting data in the tail. However, the ML and LS are not provided a good fit in the central portion of the data. The fitted density obtained from the robust estimator based on Tukey's $\rho$ function reflects better fit than the ML and LS fits in the central portion of the data. In particularly, the model obtained from the OBR estimates performs fairly well to describe the central part of the data set. The fitted densities obtained from ML, LS estimates don't seem catch $C_{max}$, the pick of the data. Therefore these estimators can not give reasonable estimate for $T_{max}$, the time taken to reach the maximum concentration. To sum up, we can clearly observe that the OBR estimation method would be preferable to fit this data set among the others.

\section{Conclusion}
\label{sec:9}
Two objectives have been considered in this study. First we have proposed to use the OBR estimation method to estimate the parameters of the MOEBXII distribution proposed by Al-Saiari et al. (2014) \cite{MOEB} with the advantage of flexibility to fit the data sets with various shapes. Second, we have considered the modeling the data sets from pharmacokinetics studies represents the changes in plasma concentrations of drugs with the MOEBXII distribution. When the estimation problem is addressed, from both simulation study and a real data example we observe that the OBR estimator exhibits strong robustness in presence of observations discordant with the assumed model. These results show that not only the OBR estimate achieves smaller RMSE for the small sample sizes but also its RMSE is smaller for the outlier cases for each sample sizes than that of the ML, LS and robust M estimators. The results of the ML and LS estimators for the outlier cases are quite different from the cases without outlier. The most striking difference is in the RMSEs of ML and LS estimates versus that of the robust estimates, especially OBR estimates. A general inspection of the table shows that a comparison of the OBR with the ML, LS and robust M estimation methods reveals the superiority of the new estimate in the outlier case and/or small sample case. When we consider the other topic, modeling pharmacokinetics data set with the MOEBXII distribution, from the real data example we can observe that the MOEBXII distribution with the OBR estimates can be a good choice for modeling the changes in plasma concentrations of drugs which is an important pharmacokinetics variable. Because estimating the parameters with the OBR estimation method would be more reliable in estimating other variables such as $C_{max}$ and $T_{max}$ other pharmacokinetics variables.

%\bibliographystyle{tfs}
%\bibliography{ref}

\begin{table}
%\sf\centering
%\tbl{The Bias and RMSE (Parenthesis) for n=25.}
%\begin{tabular}{\textwidth}{@{\extracolsep{\fill}}lllll}
\caption{The Bias and RMSE (Parenthesis) for n=25 \label{T1}}
%{
\begin{tabular}{lllll}
\hline%\toprule
Parameter $\alpha$ &&&&\\
\hline%\midrule		
 &	ML	&LS&	M (Tukey)	&OBR\\
\hline%\midrule
(3,1,1)&	0.1081 (0.2221)&	-0.4797 (0.6033)&	-0.0864 (0.0623)&	 -0.0845 \textbf{(0.0464)}\\
(3,1,2)&	0.0118 (0.2294)&	0.3000 (0.3205)&	-0.0938 (0.1003)&	 0.0069 \textbf{(0.0165)}\\
(3,2,1)&	0.1761 (0.2195)&	-0.0871 (0.4903)&	-0.0431 (0.2034)&	 -0.0011 \textbf{(0.0032)}\\
(3,2,2)&	0.1778 (0.2142)&	0.1685 (0.2696)&	0.1198 (0.1808)&	 0.0100 \textbf{(0.0036)}\\
(3,3,3)&	0.0525 (0.1987)&	0.1937 (0.1947)&	-0.0778 (0.1537)&	 0.0101 \textbf{(0.0039)}\\
(5,1,1)&	0.0328 (0.2173)&	-0.0641 (0.6265)&	-0.0990 (0.2846)&	 -0.0495 \textbf{(0.1965)}\\
(5,1,2)&	-0.0668 (0.1981)&	-0.4701 (0.1622)&	-0.4683 (0.5792)&	 -0.0102 \textbf{(0.1415)}\\
(5,2,1)&	-0.0606 (0.2083)&	-0.4757 (0.5291)&	0.0095 (0.1544)&	 0.0088 \textbf{(0.0039)}\\
(5,2,2)&	-0.0269 (0.2111)&	-0.0653 (0.7695)&	-0.4989 (0.6140)&	 0.0087 \textbf{(0.0193)}\\
\hline%\midrule
Parameter c &&&&	\\	
\hline%\midrule		
	&ML&	LS&	M (Tukey)&	OBR\\
(3,1,1)&	0.1229 (0.1026)&	0.3267 (0.2359)&	0.3239 (0.2059)&	 0.0690 \textbf{(0.0841)}\\
(3,1,2)&	0.1073 (0.0555)&	0.2684 (0.1663)&	0.2711 (0.1606)&	 -0.0005 \textbf{(0.0150)}\\
(3,2,1)&	0.1180 (0.1370)&	0.4928 (0.4435)&	0.4513 (0.3848)&	 -0.0012 \textbf{(0.0039)}\\
(3,2,2)&	0.0801 (0.1123)&	0.3417 (0.3475)&	0.3760 (0.3220)&	 -0.0011 \textbf{(0.0001)}\\
(3,3,3)&	0.0198 (0.1421)&	0.3733 (0.4870)&	0.3993 (0.4264)&	 -0.0034 \textbf{(0.0003)}\\
(5,1,1)&	0.0645 (0.0888)&	0.3310 (0.3067)&	0.4119 (0.3607)&	 0.1541 \textbf{(0.0271)}\\
(5,1,2)&	0.0947 (0.0531)&	0.3728 (0.2557)&	0.2709 (0.1452)&	 0.0010 \textbf{(0.0007)}\\
(5,2,1)&	0.0333 (0.1558)&	0.2107 (0.5419)&	0.1893 (0.3722)&	 -0.0028 \textbf{(0.0002)}\\
(5,2,2)&	0.1550 (0.1154)&	0.1436 (0.5794)&	0.4771 (0.3996)&	 -0.0016 \textbf{(0.0004)}\\
\hline%\midrule
Parameter k &&&& \\		
\hline%\midrule		
	&ML&	LS&	M (Tukey)&	OBR\\
\hline
(3,1,1)&	-0.0403 (0.0739)&	0.4731 (0.3267)&	-0.1742 (0.1155)&	 -0.0320 \textbf{(0.0635)}\\
(3,1,2)&	-0.0305 (0.1011)&	-0.3271 (0.4601)&	-0.1788 (0.1281)&	 0.0026 \textbf{(0.0020)}\\
(3,2,1)&	-0.0068 (0.0497)&	0.2188 (0.3604)&	-0.1080 (0.1292)&	 0.0006 \textbf{(0.0023)}\\
(3,2,2)&	0.0312 (0.0941)&	-0.2834 (0.3071)&	-0.2902 (0.2239)&	 0.0025 \textbf{(0.0037)}\\
(3,3,3)&	-0.0025 (0.1321)&	-0.1886 (0.2640)&	-0.0999 (0.2113)&	 0.0045 \textbf{(0.0709)}\\
(5,1,1)&	-0.2456 (0.1743)&	-0.3374 (0.4375)&	-0.3855 (0.3544)&	 0.0067 \textbf{(0.0570)}\\
(5,1,2)&	-0.0290 (0.0966)&	-0.9483 (0.9097)&	-0.9361 (0.8912)&	 -0.0017 \textbf{(0.0078)}\\
(5,2,1)&	0.0013 (0.0479)&	-0.5580 (0.3312)&	-0.5309 (0.2920)&	 0.0018 \textbf{(0.0012)}\\
(5,2,2)&	-0.0521 (0.0893)&	-0.0616 (0.1308)&	-0.0527 (0.1543)&	 0.0020 \textbf{(0.0009)}\\
\hline%\bottomrule
\end{tabular}%}
%\lipsum
%\label{T1}
\end{table}

%\clearpage
				
\begin{table}
%\sf\centering
\caption{The Bias and RMSE (Parenthesis) for n=50 \label{T2}}
%\tbl{The Bias and RMSE (Parenthesis) for n=50.}
%{\begin{tabular}{\textwidth}{@{\extracolsep{\fill}}lllll}
%{
\begin{tabular}{lllll}
\hline%\toprule
Parameter $\alpha$ &&&&\\
\hline%\midrule						
	&ML&	LS&	M (Tukey)&	OBR\\
(3,1,1)&	0.0988 (0.2140)&	-0.3154 (0.1141)&	-0.0477 (0.0240)&	 -0.0127 \textbf{(0.0047)}\\
(3,1,2)&	-0.0183 (0.2215)&	-0.3803 (0.1631)&	-0.1934 (0.0724)&	 -0.0066 \textbf{(0.0628)}\\
(3,2,1)&	0.0435 \textbf{(0.0188)}&	-0.2936 (0.1646)&	-0.0262 (0.0495)&	-0.0442 (0.0950)\\
(3,2,2)&	-0.0083 (0.2069)&	0.0970 (0.1246)&	-0.1616 (0.0687)&	 0.0156 \textbf{(0.0666)}\\
(3,3,3)&	-0.0537 \textbf{(0.0210)}&	-0.0713 (0.1523)&	-0.0840 (0.0774)&	0.1114 (0.3523)\\
(5,1,1)&	-0.0260 (0.2289)&	-0.4694 (0.2311)&	-0.4008 (0.2362)&	 -0.0178 \textbf{(0.0745)}\\
(5,1,2)&	-0.0579 (0.2139)&	-0.1342 (0.2093)&	-0.4957 (0.2457)&	 -0.0263 \textbf{(0.0214)}\\
(5,2,1)&	-0.0264 (0.2139)&	0.0147 (0.2217)&	-0.4142 (0.2188)&	 -0.0259 \textbf{(0.0013)}\\
(5,2,2)&	-0.0481 (0.1915)&	0.0584 (0.2048)&	-0.4999 (0.2500)&	 -0.0331 \textbf{(0.0727)}\\
\hline%\midrule	
Parameter c &&&&\\	
\hline%\midrule				
	&ML	&LS	&M (Tukey)	&OBR\\
\hline%\midrule	
(3,1,1)&	0.0466 (0.0479)&	0.2047 (0.0652)&	0.1930 (0.0645)&	 0.0015 \textbf{(0.0184)}\\
(3,1,2)&	0.0267 (0.0423)&	0.1867 (0.0529)&	0.1461 (0.0529)&	 0.0011 \textbf{(0.0096)}\\
(3,2,1)&	0.0477 (0.0710)&	0.3217 (0.1453)&	0.2983 (0.1346)&	 0.0054 \textbf{(0.0512)}\\
(3,2,2)&	0.0501 (0.0359)&	0.3102 (0.1353)&	0.2588 (0.1226)&	 -0.0024 \textbf{(0.0021)}\\
(3,3,3)&	0.0307 (0.0582)&	0.3692 (0.1865)&	0.3227 (0.1811)&	 -0.0287 \textbf{(0.0215)}\\
(5,1,1)&	0.0505 \textbf{(0.0342)}&	0.1857 (0.0528)&	0.2200 (0.0804)&	0.0016 (0.0469)\\
(5,1,2)&	0.0227 \textbf{(0.0146)}&	0.1026 (0.0547)&	0.3426 (0.1504)&	0.0012 (0.0648)\\
(5,2,1)&	0.0592 (0.0816)&	0.1588 (0.1661)&	0.3253 (0.1614)&	 0.0043 \textbf{(0.0387)}\\
(5,2,2)&	0.0613 (0.0893)&	0.1208 (0.0954)&	0.4289 (0.2160)&	 0.0024 \textbf{(0.0342)}\\
\hline%\midrule	
Parameter k &&&&\\				
\hline%\midrule	
	&ML	&LS&	M (Tukey)	&OBR\\
\hline%\midrule	
(3,1,1)&	-0.0218 (0.0198)&	-0.0982 (0.0281)&	-0.2219 (0.0605)&	 -0.0026 \textbf{(0.0179)}\\
(3,1,2)&	0.0469 (0.0505)&	0.1787 (0.0472)&	-0.1068 (0.0315)&	 -0.0021 \textbf{(0.0059)}\\
(3,2,1)&	0.0032 (0.0659)&	-0.0263 (0.0565)&	-0.0805 \textbf{(0.0558)}&	-0.0049 (0.0613)\\
(3,2,2)&	0.0104 \textbf{(0.0461)}&	-0.2862 (0.1539)&	-0.1492 (0.0642)&	0.0069 (0.0086)\\
(3,3,3)&	-0.0239 (0.0933)&	-0.1003 (0.1198)&	-0.2754 (0.1172)&	 0.0456 \textbf{(0.0650)}\\
(5,1,1)&	-0.0299 (0.0239)&	-0.4980 (0.2481)&	-0.4967 (0.2468)&	 -0.0023 \textbf{(0.0012)}\\
(5,1,2)&	-0.0134 \textbf{(0.0330)}&	-0.0235 (0.0946)&	-0.4470 (0.2499)&	-0.0049 (0.0771)\\
(5,2,1)&	-0.0006 \textbf{(0.0210)}&	0.0010 (0.0487)&	-0.4959 (0.2462)&	-0.0033 (0.0246)\\
(5,2,2)&	-0.0033 \textbf{(0.0242)}&	0.0197 (0.0909)&	-0.4989 (0.2489)&	-0.0051 (0.0279)\\
\hline%\bottomrule
\end{tabular}%}
%\label{T2}
\end{table}
%\clearpage
				
\begin{table}
%\sf\centering
\caption{The Bias and RMSE (Parenthesis) for n=100 \label{T3}}
%\tbl{The Bias and RMSE (Parenthesis) for n=100.}
%{
\begin{tabular}{lllll}
\hline%\toprule				
Parameter $\alpha$	&&&&\\	
\hline%\midrule			
	&ML	&LS	&M (Tukey)	&OBR\\
\hline%\midrule	
(3,1,1)&	0.0097 \textbf{(0.0355)}&	-0.2899 (0.1072)&	-0.0607 (0.0397)&	-0.1376 (0.2175)\\
(3,1,2)&	-0.0191 \textbf{(0.0368)}&	-0.3663 (0.1805)&	-0.2242 (0.0980)&	0.0879 (0.2121)\\
(3,2,1)&	0.0325 \textbf{(0.0347)}&	-0.3331 (0.1830)&	-0.0721 (0.0711)&	0.0540 (0.1974)\\
(3,2,2)&	0.0249 \textbf{(0.0365)}&	0.0373 (0.1226)&	-0.1339 (0.0878)&	-0.0457 (0.2264)\\
(3,3,3)&	-0.0006 \textbf{(0.0328)}&	-0.1143 (0.1465)&	-0.1815 (0.1187)&	-0.0452 (0.2125)\\
(5,1,1)&	-0.0487 \textbf{(0.0357)}&	-0.4458 (0.2158)&	-0.3646 (0.2165)&	-0.0300 (0.2173)\\
(5,1,2)&	-0.0168 \textbf{(0.0349)}&	-0.1446 (0.1883)&	-0.4709 (0.2435)&	-0.0504 (0.2126)\\
(5,2,1)&	0.0293 \textbf{(0.0354)}&	0.0118 (0.2107)&	-0.3618 (0.2227)&	0.0597 (0.2211)\\
(5,2,2)&	0.0137 \textbf{(0.0368)}&	-0.0741 (0.2187)&	-0.4960 (0.2468)&	0.0701 (0.1997)\\
\hline%\midrule	
Parameter c	&&&&\\
\hline%\midrule				
	&ML	&LS	&M (Tukey)	&OBR\\
\hline%\midrule	
(3,1,1)&	0.0315 \textbf{(0.0164)}&	0.2066 (0.0980)&	0.2044 (0.0986)&	0.0557 (0.0492)\\
(3,1,2)&	0.0254 \textbf{(0.0134)}&	0.1471 (0.0566)&	0.1561 (0.0598)&	0.0115 (0.0177)\\
(3,2,1)&	-0.0001 \textbf{(0.0282)}&	0.2233 (0.1595)&	0.2116 (0.1534)&	0.0666 (0.1013)\\
(3,2,2)&	0.0034 \textbf{(0.0203)}&	0.3162 (0.1652)&	0.3160 (0.1594)&	0.1252 (0.0733)\\
(3,3,3)&	0.0232 \textbf{(0.0288)}&	0.3041 (0.1628)&	0.2714 (0.1573)&	0.1005 (0.1030)\\
(5,1,1)&	0.2304 (0.0941)&	0.1675 (0.0636)&	0.0613 (0.0446)&	 0.0152 \textbf{(0.0211)}\\
(5,1,2)&	0.0318 \textbf{(0.0154)}&	0.0833 (0.0476)&	0.3125 (0.1322)&	0.0323 (0.0344)\\
(5,2,1)&	0.3095 \textbf{(0.1688)}&	0.0565 (0.1553)&	0.0231 (0.0306)&	0.1173 (0.1171)\\
(5,2,2)&	0.0069 \textbf{(0.0290)}&	0.0375 (0.0949)&	0.3916 (0.1986)&	0.0320 (0.0849)\\
\hline%\midrule	
Parameter k&&&&\\
\hline%\midrule					
	&ML	&LS	&M (Tukey)	&OBR\\
\hline%\midrule	
(3,1,1)&	0.0071 (0.0218)&	-0.0794 (0.0507)&	-0.2181 (0.0733)&	 -0.0462 \textbf{(0.0118})\\
(3,1,2)&	-0.0088 (0.0514)&	0.1005 (0.0536)&	-0.1030 (0.0447)&	 -0.0233 \textbf{(0.0287)}\\
(3,2,1)&	0.0049 \textbf{(0.0128)}&	0.0109 (0.0759)&	-0.1993 (0.0722)&	-0.0009 (0.0382)\\
(3,2,2)&	-0.2288 (0.0843)&	-0.2906 (0.1341)&	-0.0071 \textbf{(0.0223)}&	-0.0313 (0.0732)\\
(3,3,3)&	-0.0237 \textbf{(0.0302)}&	-0.1285 (0.1235)&	-0.1397 (0.1128)&	-0.0007 (0.0958)\\
(5,1,1)&	0.0130 \textbf{(0.0226)}&	-0.4924 (0.2428)&	-0.4952 (0.2455)&	-0.0218 (0.0363)\\
(5,1,2)&	-0.0045 \textbf{(0.0230)}&	-0.0171 (0.1301)&	-0.4933 (0.2456)&	-0.0490 (0.0512)\\
(5,2,1)&	0.0153 \textbf{(0.0106)}&	0.0231 (0.0478)&	-0.4908 (0.2415)&	-0.0143 (0.0306)\\
(5,2,2)&	0.0152 \textbf{(0.0215)}&	-0.0074 (0.0779)&	-0.4995 (0.2495)&	0.0207 (0.0566)\\
\hline%\bottomrule
\end{tabular}%}
%\label{T3}
\end{table}

\begin{table}
%\sf\centering
\caption{The Bias and RMSE (Parenthesis) for n=25 with one outlier \label{T4}}
%\tbl{The Bias and RMSE (Parenthesis) for n=25 with one outlier.}
%{
\begin{tabular}{lllll}
\hline%\toprule				
Parameter $\alpha$ &&&&\\ 	
\hline			
	&ML	&LS	&M (Tukey)	&OBR\\
(3,1,1)&	0.5260 (0.7583)&	0.3937 (0.6005)&	0.0648 (0.0974)&	 0.0675 \textbf{(0.0334)}\\
(3,1,2)&	0.7677 (1.0733)&	-0.2801 (0.9506)&	0.2737 (0.3186)&	 0.1322 \textbf{(0.1359)}\\
(3,2,1)&	0.6696 (0.8518)&	0.7122 (0.5237)&	0.0598 (0.0664)&	 0.0693 \textbf{(0.0384)}\\
(3,2,2)&	0.9541 (0.9890)&	-0.3073 (0.2022)&	0.2807 (0.5148)&	 0.1792 \textbf{(0.1211)}\\
(3,3,3)&	0.3557 (0.1608)&	0.8745 (0.8261)&	0.1740 (0.2903)&	 0.0327 \textbf{(0.0782)}\\
(5,1,1)&	0.7769 (0.8408)&	0.6287 (0.5721)&	0.3778 (0.2233)&	 0.1630 \textbf{(0.0783)}\\
(5,1,2)&	0.8666 (0.9375)&	0.7299 (0.7325)&	0.4934 (0.2456)&	 0.0133 \textbf{(0.0215)}\\
(5,2,1)&	0.6980 (0.9422)&	0.5848 (0.5159)&	0.3630 (0.2123)&	 -0.0127 \textbf{(0.0257)}\\
(5,2,2)&	0.9814 (0.9669)&	0.8705 (0.8508)&	0.4707 (0.7286)&	 0.2630 \textbf{(0.2471)}\\
\hline%\midrule
Parameter c	&&&&\\			
\hline%\midrule
	&ML	&LS	&M (Tukey)	&OBR\\
\hline%\midrule
(3,1,1)&	-0.2184 (0.1522)&	-0.3247 (0.2277)&	-0.2399 (0.1012)&	 -0.0170 \textbf{(0.0016)}\\
(3,1,2)&	-0.1275 (0.0787)&	-0.3053 (0.2022)&	-0.2481 (0.1208)&	 -0.0298 \textbf{(0.0158)}\\
(3,2,1)&	-0.2308 (0.3010)&	-0.4433 (0.4108)&	-0.2805 (0.1529)&	 -0.0303 \textbf{(0.0039)}\\
(3,2,2)&	0.0665 (0.1365)&	-0.2382 (0.2869)&	-0.1332 (0.1106)&	 -0.0491 \textbf{(0.0328)}\\
(3,3,3)&	0.4780 (0.3820)&	-0.2563 (0.3921)&	-0.4035 (0.3110)&	 -0.1655 \textbf{(0.1587)}\\
(5,1,1)&	-0.2185 (0.2043)&	-0.4727 (0.3573)&	-0.2399 (0.1081)&	 -0.0214 \textbf{(0.0015)}\\
(5,1,2)&	-0.0734 (0.1097)&	-0.3024 (0.1942)&	-0.3290 (0.1483)&	 0.0227 \textbf{(0.0528)}\\
(5,2,1)&	-0.2499 (0.4091)&	-0.5456 (0.5562)&	-0.1779 (0.1339)&	 -0.0251 \textbf{(0.0810)}\\
(5,2,2)&	-0.0615 (0.2139)&	-0.5764 (0.5222)&	-0.3317 (0.1961)&	 -0.0728 \textbf{(0.0270)}\\
\hline%\midrule
Parameter k&&&&\\
\hline%\midrule				
	&ML	&LS	&M (Tukey)	&OBR\\
\hline%\midrule
(3,1,1)&	0.2541 (0.1131)&	-0.3756 (0.3442)&	0.3033 (0.1241)&	 0.0213 \textbf{(0.0048)}\\
(3,1,2)&	0.4746 (0.3717)&	0.4407 (0.5491)&	0.2398 (0.1200)&	 0.0637 \textbf{(0.0899)}\\
(3,2,1)&	0.2620 (0.1153)&	-0.5760 (0.4492)&	0.2638 (0.1155)&	 0.0170 \textbf{(0.0019)}\\
(3,2,2)&	0.5505 (0.3801)&	0.5809 (0.3841)&	0.2605 (0.1061)&	 0.0583 \textbf{(0.0551)}\\
(3,3,3)&	0.9466 (0.9067)&	0.7017 (0.5734)&	0.4727 (0.4533)&	 0.2293 \textbf{(0.1271)}\\
(5,1,1)&	0.2190 (0.1047)&	0.6239 (0.4033)&	0.4945 (0.2448)&	 0.0280 \textbf{(0.0026)}\\
(5,1,2)&	0.2445 (0.3798)&	0.9613 (0.9299)&	0.4976 (0.9480)&	 0.2095 \textbf{(0.2479)}\\
(5,2,1)&	0.1730 (0.1975)&	0.5925 (0.3678)&	0.4706 (0.2271)&	 -0.0004 \textbf{(0.0538)}\\
(5,2,2)&	0.4730 (0.3290)&	0.9820 (0.9687)&	0.4554 (0.2455)&	 0.0791 \textbf{(0.0385)}\\
\hline%\bottomrule
\end{tabular}%}
%\label{T4}
\end{table}

\clearpage
%Table 5 The Bias and RMSE (Parenthesis) for n=50 with two outliers				
\begin{table}
%\sf\centering
\caption{The Bias and RMSE (Parenthesis) for n=50 with one outlier \label{T5}}
%\tbl{The Bias and RMSE (Parenthesis) for n=50 with one outlier.}
%{
\begin{tabular}{lllll}
\hline%\toprule				
Parameter $\alpha$&&&&\\
\hline%\midrule				
	&ML	&LS	&M (Tukey)	&OBR\\
\hline%\midrule
(3,1,1)&	0.7658 (0.7925)&	0.2757 (0.5942)&	0.0844 (0.0336)&	 0.0972 \textbf{(0.0192)}\\
(3,1,2)&	0.9122 (0.9377)&	-0.2170 (0.1263)&	0.2961 (0.1128)&	 0.0691 \textbf{(0.0969)}\\
(3,2,1)&	0.8448 (0.8977)&	0.7232 (0.5303)&	0.0696 (0.0452)&	 0.0194 \textbf{(0.0203)}\\
(3,2,2)&	0.9860 (0.9818)&	-0.2744 (0.1613)&	0.1150 (0.1290)&	 0.3149 \textbf{(0.0935)}\\
(3,3,3)&	0.9464 (0.9216)&	0.2705 (0.2294)&	0.3848 (0.1637)&	 -0.0039 \textbf{(0.0482)}\\
(5,1,1)&	0.5430 (0.8791)&	0.6820 (0.5746)&	0.3746 (0.2256)&	 0.1248 \textbf{(0.0341)}\\
(5,1,2)&	0.8018 (0.9470)&	0.9043 (0.8835)&	0.4865 (0.2431)&	 0.1075 \textbf{(0.2400)}\\
(5,2,1)&	0.8957 (0.9413)&	0.6427 (0.5411)&	0.4286 (0.2280)&	 0.0585 \textbf{(0.1521)}\\
(5,2,2)&	0.9518 (0.9853)&	0.9194 (0.8797)&	0.4780 (0.2481)&	 0.1677 \textbf{(0.2322)}\\
\hline%\midrule
Parameter c&&&&\\
\hline%\midrule				
	&ML	&LS	&M (Tukey)	&OBR\\
\hline%\midrule
(3,1,1)&	-0.1187 (0.1086)&	-0.2660 (0.1783)&	-0.1984 (0.0917)&	 -0.0161 \textbf{(0.0004)}\\
(3,1,2)&	-0.0095 (0.0199)&	-0.1356 (0.0502)&	-0.1181 (0.0473)&	 -0.0036 \textbf{(0.0008)}\\
(3,2,1)&	-0.1724 (0.2018)&	-0.2910 (0.2963)&	-0.1612 (0.1244)&	 -0.0281 \textbf{(0.0024)}\\
(3,2,2)&	-0.1955 (0.1236)&	-0.2882 (0.2294)&	0.0101 (0.0581)&	 -0.0290 \textbf{(0.0045)}\\
(3,3,3)&	0.5491 (0.3680)&	-0.2338 (0.2683)&	-0.0770 (0.1276)&	 -0.0786 \textbf{(0.0182)}\\
(5,1,1)&	-0.1099 (0.0960)&	-0.3937 (0.2707)&	-0.2269 (0.1000)&	 -0.0159 \textbf{(0.0004)}\\
(5,1,2)&	-0.0081 (0.0340)&	-0.3363 (0.1934)&	-0.3118 (0.1338)&	 -0.0102 \textbf{(0.0010)}\\
(5,2,1)&	-0.1651 (0.2883)&	-0.5914 (0.5327)&	-0.3109 (0.1790)&	 -0.0341 \textbf{(0.0034)}\\
(5,2,2)&	0.1245 (0.1309)&	-0.5378 (0.4875)&	-0.3568 (0.1874)&	 -0.0336 \textbf{(0.0051)}\\
\hline%\midrule
Parameter k	&&&&\\
\hline%\midrule			
	&ML	&LS	&M (Tukey)	&OBR\\
\hline%\midrule
(3,1,1)&	0.2419 (0.1065)&	-0.3498 (0.3793)&	0.2529 (0.0928)&	 0.0223 \textbf{(0.0009)}\\
(3,1,2)&	0.5264 (0.3272)&	0.5314 (0.4231)&	0.2287 (0.0853)&	 0.0217 \textbf{(0.0074)}\\
(3,2,1)&	0.2877 (0.1012)&	-0.5686 (0.3724)&	0.2880 (0.1090)&	 0.0161 \textbf{(0.0011)}\\
(3,2,2)&	0.6017 (0.4020)&	0.6141 (0.4017)&	0.2857 (0.1095)&	 0.0346 \textbf{(0.0075)}\\
(3,3,3)&	0.9529 (0.9132)&	0.6530 (0.4790)&	0.2459 (0.1222)&	 0.0912 \textbf{(0.0250)}\\
(5,1,1)&	0.1709 (0.0692)&	0.6144 (0.3888)&	0.4969 (0.2470)&	 0.0198 \textbf{(0.0008)}\\
(5,1,2)&	0.3460 (0.1854)&	0.9919 (0.9857)&	0.4437 (0.2444)&	 0.0243 \textbf{(0.0083)}\\
(5,2,1)&	0.2067 (0.0796)&	0.6092 (0.3788)&	0.4968 (0.2470)&	 0.0118 \textbf{(0.0081)}\\
(5,2,2)&	0.4108 (0.2310)&	0.9936 (0.9879)&	0.4991 (0.2491)&	 0.0377 \textbf{(0.0080)}\\
\hline%\bottomrule
\end{tabular}%}
%\label{T5}
\end{table}

\clearpage
%Table 6 The Bias and RMSE (Parenthesis) for n=100 with four outliers				
\begin{table}
%\sf\centering
\caption{The Bias and RMSE (Parenthesis) for n=100 with one outlier \label{T6}}
%\tbl{The Bias and RMSE (Parenthesis) for n=100 with one outlier.}
%{
\begin{tabular}{lllll}
\hline%\toprule		
Parameter $\alpha$&&&&\\
\hline%\midrule				
	&ML	&LS	&M (Tukey)	&OBR\\
\hline%\midrule
(3,1,1)&	0.8537 (0.9116)&	0.7543 (0.6559)&	0.0566 (0.0316)&	 0.0104 \textbf{(0.0018)}\\
(3,1,2)&	0.9586 (0.9905)&	-0.3791 (0.2258)&	0.2690 (0.1110)&	 0.0130 \textbf{(0.0016)}\\
(3,2,1)&	0.9976 (0.9954)&	0.8250 (0.6824)&	0.0300 (0.0309)&	 0.0137 \textbf{(0.0290)}\\
(3,2,2)&	0.4103 (0.1894)&	-0.2510 (0.1249)&	0.3399 (0.1359)&	 0.0192 \textbf{(0.0018)}\\
(3,3,3)&	0.9669 (0.9403)&	-0.2345 (0.1484)&	0.4405 (0.1998)&	 0.0393 \textbf{(0.0753)}\\
(5,1,1)&	0.9206 (0.9225)&	0.8001 (0.8085)&	0.4078 (0.2375)&	 0.0153 \textbf{(0.0003)}\\
(5,1,2)&	0.9836 (0.9710)&	0.9266 (0.8986)&	0.4604 (0.2496)&	 0.0318 \textbf{(0.0386)}\\
(5,2,1)&	0.9471 (1.0099)&	0.7665 (0.6802)&	0.4633 (0.2418)&	 0.0180 \textbf{(0.0407)}\\
(5,2,2)&	0.4346 (0.2033)&	0.9036 (0.8763)&	0.4800 (0.2500)&	 0.0268 \textbf{(0.0034)}\\
\hline%\midrule
Parameter c&&&&\\
\hline%\midrule				
	&ML	&LS	&M (Tukey)	&OBR\\
\hline%\midrule
(3,1,1)&	-0.0742 (0.4301)&	-0.1888 (0.6320)&	-0.1622 (0.0572)&	 -0.0013 \textbf{(0.0231)}\\
(3,1,2)&	-0.0144 (0.0093)&	-0.1600 (0.4116)&	-0.1358 (0.0390)&	 -0.0015 \textbf{(0.0012)}\\
(3,2,1)&	-0.2041 (0.1376)&	-0.4140 (0.3012)&	-0.2573 (0.1460)&	 -0.0035 \textbf{(0.0155)}\\
(3,2,2)&	0.0852 (0.0584)&	-0.2921 (0.1768)&	-0.1749 (0.0913)&	 -0.0036 \textbf{(0.0451)}\\
(3,3,3)&	0.5257 (0.3139)&	-0.3673 (0.3133)&	-0.1820 (0.1206)&	 -0.0102 \textbf{(0.0600)}\\
(5,1,1)&	-0.0490 (0.0431)&	-0.3632 (0.2194)&	-0.2476 (0.1015)&	 -0.0014 \textbf{(0.0245)}\\
(5,1,2)&	0.0414 (0.0126)&	-0.2780 (0.1118)&	-0.2837 (0.1087)&	 -0.0020 \textbf{(0.0012)}\\
(5,2,1)&	-0.1072 (0.1860)&	-0.6308 (0.5946)&	-0.3055 (0.1529)&	 -0.0031 \textbf{(0.0117)}\\
(5,2,2)&	0.3992 (0.0753)&	-0.5101 (0.3795)&	-0.3966 (0.1878)&	 -0.0032 \textbf{(0.0343)}\\
\hline%\midrule
Parameter k&&&&\\
\hline%\midrule				
	&ML	&LS	&M (Tukey)	&OBR\\
\hline%\midrule
(3,1,1)&	0.2783 (0.0949)&	-0.9356 (1.0459)&	0.2708 (0.0819)&	 0.0020 \textbf{(0.0576)}\\
(3,1,2)&	0.5343 (0.3174)&	0.6186 (0.4029)&	0.2308 (0.1867)&	 0.0040 \textbf{(0.0915)}\\
(3,2,1)&	0.3315 (0.1188)&	-1.1387 (1.3363)&	0.3016 (0.0988)&	 0.0025 \textbf{(0.0857)}\\
(3,2,2)&	0.5936 (0.3715)&	0.5878 (0.3550)&	0.2493 (0.1933)&	 0.0050 \textbf{(0.1096)}\\
(3,3,3)&	0.9831 (0.9681)&	0.8888 (0.8074)&	0.2834 (0.1353)&	 0.0124 \textbf{(0.0084)}\\
(5,1,1)&	0.1908 (0.0588)&	0.6082 (0.3774)&	0.4993 (0.2493)&	 0.0019 \textbf{(0.0458)}\\
(5,1,2)&	0.3245 (0.1361)&	0.9955 (0.9914)&	0.4971 (0.2472)&	 0.0055 \textbf{(0.0107)}\\
(5,2,1)&	0.2006 (0.0638)&	0.6048 (0.3712)&	0.4987 (0.2488)&	 0.0021 \textbf{(0.0549)}\\
(5,2,2)&	0.3428 (0.1376)&	0.9992 (0.9985)&	0.3428 (0.1376)&	 0.0041 \textbf{(0.0679)}\\
\hline%\bottomrule
\end{tabular}%}
%\label{T6}
\end{table}

\clearpage
%Table 7 The Bias and RMSE (Parenthesis) for n=50 with four outliers				
\begin{table}
%\sf\centering
\caption{The Bias and RMSE (Parenthesis) for n=50 with four outlier \label{T7}}
%\tbl{The Bias and RMSE (Parenthesis) for n=50 with four outlier.}
%{
\begin{tabular}{lllll}
\hline%\toprule				
Parameter $\alpha$\\
\hline%\midrule				
	&ML	&LS	&M (Tukey)	&OBR\\
\hline%\midrule
(3,1,1)&	1.8192 (1.6599)&	0.8780 (0.7723)&	0.1795 (0.0634)&	 0.0834 \textbf{(0.0116)}\\
(3,1,2)&	2.3995 (2.9045)&	0.4393 (0.2599)&	0.4573 (0.2555)&	 0.1753 \textbf{(0.0405)}\\
(3,2,1)&	2.1213 (2.6922)&	0.8396 (0.7130)&	0.2445 (0.1116)&	 0.1012 \textbf{(0.0123)}\\
(3,2,2)&	2.6898 (2.2763)&	0.3345 (0.1911)&	0.4635 (0.2822)&	 0.1896 \textbf{(0.0679)}\\
(3,3,3)&	2.9292 (2.5814)&	0.3273 (0.1249)&	0.5968 (0.3730)&	 0.2891 \textbf{(0.1190)}\\
(5,1,1)&	2.9800 (3.0851)&	0.8503 (0.7336)&	0.2737 (0.1275)&	 0.1125 \textbf{(0.0160)}\\
(5,1,2)&	3.9945 (4.3873)&	0.9993 (0.9986)&	0.9991 (0.9982)&	 0.2696 \textbf{(0.1125)}\\
(5,2,1)&	3.7069 (3.1097)&	0.9545 (0.9885)&	0.3372 (0.1666)&	 0.1335 \textbf{(0.0220)}\\
(5,2,2)&	3.5945 (4.1334)&	0.9853 (0.9761)&	0.9940 (0.9898)&	 0.2896 \textbf{(0.1202)}\\
\hline%\midrule
Parameter c	&&&&\\
\hline%\midrule			
	&ML	&LS	&M (Tukey)	&OBR\\
\hline%\midrule
(3,1,1)&	0.2709 (0.1274)&	0.2205 (0.0778)&	0.2232 (0.0878)&	 0.0150 \textbf{(0.0005)}\\
(3,1,2)&	0.3592 (0.1972)&	0.1978 (0.0636)&	0.1986 (0.0697)&	 0.0167 \textbf{(0.0004)}\\
(3,2,1)&	0.7857 (1.1572)&	0.5197 (0.6674)&	0.3933 (0.2462)&	 0.0329 \textbf{(0.0015)}\\
(3,2,2)&	0.9072 (1.1683)&	0.3664 (0.2268)&	0.3465 (0.1961)&	 0.0417 \textbf{(0.0044)}\\
(3,3,3)&	1.3173 (2.0957)&	0.3082 (0.1525)&	0.3728 (0.2224)&	 0.0800 \textbf{(0.0098)}\\
(5,1,1)&	0.3561 (0.2606)&	0.3810 (0.2948)&	0.3490 (0.2166)&	 0.0128 \textbf{(0.0002)}\\
(5,1,2)&	0.4625 (0.3055)&	0.2670 (0.1102)&	0.2925 (0.1343)&	 0.0162 \textbf{(0.0004)}\\
(5,2,1)&	0.9675 (1.4720)&	0.8075 (1.0989)&	0.5793 (0.4692)&	 0.0352 \textbf{(0.0017)}\\
(5,2,2)&	0.9994 (1.2173)&	0.4066 (0.2439)&	0.4501 (0.3097)&	 0.0402 \textbf{(0.0025)}\\
\hline%\midrule
Parameter k&&&&\\				
\hline%\midrule
	&ML	&LS	&M (Tukey)	&OBR\\
\hline%\midrule
(3,1,1)&	0.5637 (0.3516)&	1.8390 (1.6041)&	0.3359 (0.1325)&	 0.0196 \textbf{(0.0008)}\\
(3,1,2)&	1.4464 (2.1420)&	0.8282 (0.7011)&	0.3686 (0.2293)&	 0.0474 \textbf{(0.0031)}\\
(3,2,1)&	0.6755 (0.4794)&	1.4009 (2.7706)&	0.3795 (0.1661)&	 0.0210 \textbf{(0.0006)}\\
(3,2,2)&	1.6785 (2.8440)&	0.8418 (0.7197)&	0.5169 (0.3583)&	 0.0521 \textbf{(0.0067)}\\
(3,3,3)&	2.8503 (3.1275)&	0.8355 (0.7122)&	0.3288 (0.1913)&	 0.0926 \textbf{(0.0129)}\\
(5,1,1)&	0.5399 (0.3318)&	0.6267 (0.4037)&	0.6583 (0.4414)&	 0.0159 \textbf{(0.0003)}\\
(5,1,2)&	1.4151 (2.0741)&	0.9999 (0.9998)&	0.7133 (0.7656)&	 0.0454 \textbf{(0.0032)}\\
(5,2,1)&	0.6832 (0.4851)&	0.6485 (0.4277)&	0.6753 (0.4605)&	 0.0205 \textbf{(0.0006)}\\
(5,2,2)&	1.6970 (1.8872)&	0.9983 (0.9967)&	0.1379 (0.1287)&	 0.0497 \textbf{(0.0038)}\\
\hline%\bottomrule
\end{tabular}%}
%\label{T7}
\end{table}

%Table 8 ML, LS, M and OBR Estimates for ibuprofen data			
\begin{table}
%\sf\centering
\caption{ML, LS, M and OBR Estimates for ibuprofen data	.\label{T8}}
%\tbl{ML, LS, M and OBR Estimates for ibuprofen data.}
%{
\begin{tabular}{llll}
\hline%\toprule
	&$\widehat{\alpha}$	&$\widehat{c}$	&$\widehat{k}$\\
\hline%\midrule
MLE&	23.7002&	1.7654&	0.9243\\
LSE&	37.002&	1.83647&	0.9726\\
M Estimation (Tukey)&	41.1842&	2.29&	0.8721\\
OBRE&	34.5757&	2.5723&	0.7726\\
\hline%\bottomrule
\end{tabular}%}
%\label{T8}
\end{table}

\end{document}